\documentclass[sigconf]{acmart}
\usepackage{natbib}
\usepackage{tabularx}
\usepackage[autostyle]{csquotes} 
\usepackage{url} 
\usepackage{todonotes}
\usepackage{float}
\usepackage{array}
\usepackage{tabu}

\acmConference[OpenSuCo 2017]{2017 Workshop on Open Source Supercomputing}{November 2017}{Denver, Colorado USA}
\acmYear{2017}
\copyrightyear{2017}

\begin{document}

\title{Toward Common Components for Open Workflow Systems}

\author{Jay Jay Billings}
\orcid{orcid.org/0000-0001-8811-2688}
\affiliation{
\institution{Oak Ridge National Laboratory and The Bredesen Center for Interdisciplinary Research and Graduate Education, The University of Tennessee - Knoxville}
\streetaddress{PO Box 2008 MS 6173}
\city{Oak Ridge}
\state{TN}
\country{USA}
\postcode{37831}}
\email{billingsjj@ornl.gov\\ Twitter: @jayjaybillings} 

\author{Shantenu Jha}
\affiliation{
\institution{Computational Science Initiative, Brookhaven National Laboratory and Rutgers University}
\city{Upton}
\state{NY}
\country{USA}
\postcode{11973}}
\email{shantenu.jha@rutgers.edu}

\begin{abstract}

The role of scalable high-performance workflows and flexible workflow
management systems that can support multiple simulations will continue to
increase in importance. For example, with the end of Dennard scaling, there is
a need to substitute a single long running simulation with multiple repeats of
shorter simulations, or concurrent replicas. Further, many scientific problems
involve ensembles of simulations in order to solve a higher-level problem or
produce statistically meaningful results. However most supercomputing software
development and performance enhancements have focused on optimizing single-
simulation performance. On the other hand, there is a strong inconsistency in
the definition and practice of workflows and workflow management systems. This
inconsistency often centers around the difference between several different
types of workflows, including modeling and simulation, grid, uncertainty
quantification, and purely conceptual workflows. This work explores this
phenomenon by examining the different types of workflows and workflow
management systems, reviewing the perspective of a large supercomputing
facility, examining the common features and problems of workflow management
systems, and finally presenting a proposed solution based on the concept of
common building blocks. The implications of the continuing proliferation of
workflow management systems and the lack of interoperability between these
systems are discussed from a practical perspective. In doing so, we have begun
an investigation of the design and implementation of open workflow systems for
supercomputers based upon common components.
\end{abstract}

\maketitle

\underline{Notice of Copyright:} This manuscript has been authored by UT-
Battelle, LLC under Contract No. DEAC05-00OR22725 with the U.S. Department of
Energy. The United States Government retains and the publisher, by accepting
the article for publication, acknowledges that the United States Government
retains a nonexclusive, paid-up, irrevocable, world-wide license to publish or
reproduce the published form of this manuscript, or allow others to do so, for
United States Government purposes. The Department of Energy will provide
public access to these results of federally sponsored research in accordance
with the DOE Public Access Plan (http://energy.gov/downloads/doe-public-access-plan).

\section{Introduction}

Suppose for a moment that there is an interesting activity that would benefit
from automation, which is known because the activity exhibits the following
properties: 
\begin{itemize} 
\item The goal of the activity is known and desirable.  
\item The tasks to achieve the goal and complete the activity are also known 
and, furthermore, highly repetitive even in cases where decisions must be 
made to continue.  
\item The results of achieving this goal can be consumed or processed in 
standard ways.  
\end{itemize}

This example may be recognized by many as a description---but not a definition---of a workflow. Experts from many backgrounds can easily think
of activities that fit this description and even systems that automate the
activity. However, each expert will probably also imagine a different workflow:
a businessperson might imagine the workflow for processing payments; a medical
professional might imagine updating medical charts and records; and
scientists might imagine performing an analysis with modeling and simulation
software, analyzing a large amount of data, or quantifying uncertainty. Within
the scientific community this has led to a rather predictable situation:
Everyone has a different definition of workflow and has created their own
systems for managing and processing workflows.

This leads to some very practical consequences for scientific
workflows. In spite of the similarities in high-level abstractions and
higher-order concepts, extremely specialized software solutions and communities
have developed to process scientific workflows. These differences hold across
scientific problems, all generally providing some level of service that was not
or perhaps is not available in a regular programming language, system library,
or problem-solving workbench. These systems have accreted workflow management
capabilities over time that have effectively resulted in the creation of large,
monolithic software stacks that cannot communicate between each other, require a
very large amount of expertise to use, often put very high demands on back-end
systems either by design or through assumptions, and are often too specialized
to jump between workflow execution for data analysis and modeling and
simulation. 

Recent developments suggest that this may be neither desirable nor necessary.
The continued scalability, sophistication, and maintainability of large, monolithic systems is called into question as scientific problems become more complex, functionality moves from libraries to operating systems, and open source development continues to rise as the dominant means of collaborating on software development. Software complexity, in particular, often makes it
impossible for development on large systems to scale to the required level 
because the accretion of new capabilities means managing larger pools of
people and a larger software development effort. One obvious alternative with
some degree of historical precedence in the field is to develop common
building blocks that provide common services used to both define and execute
workflows. Such an approach not only makes it possible to coalesce around a
standard definition and understanding of workflows, but also make it possible to separate and
distribute the work required to construct the building blocks from the effort to
define workflows and to create workflow management systems that may share the
building blocks while retaining required customizations. This article
contributes to the ongoing discussion by providing
\begin{itemize}
\item an illustration of the diverse nature of
scientific workflows (\S \ref{workflows}) that describes the different
areas where scientific workflows and systems have appeared in the literature,
how they have been classified in the past, and the arguments around coalescence
that are driven by calls of interoperability (\S\ref{interop} and \S\ref{commonFunc});
\item a description of the necessary
subset of functionality that is common across a number of scientific workflow
management systems that would, in principle, be good candidates for
consolidation and sharing (\S \ref{buildings-blocks}); and
\item an understanding of
these common elements as building blocks and how composing these building blocks
addresses a number of the problems not easily addressed by the monolithic design
of existing systems (\S\ref{buildings-blocks} and \S\ref{discussion}).
\end{itemize}

\section{The Diversity of Workflow Models}\label{workflows}

One of the most challenging aspects of studying workflows is the way the
vocabulary has been unintentionally overloaded.  It is clearer to
understand it by starting from a historical perspective.

The use and study of workflows and the initial implementation of workflow
management systems (i.e., systems that manage one or more activities related to
workflows), and especially workflow execution, was developed in the business world to addres the need to automate business processes. Lud\"{a}scher et~al.  ascribe the
origins of workflows and workflow management systems to ``office automation''
trends in the 1970s \cite{ludascher_scientific_2006}. Van~Der~Aalst argues that
``workflows'' arose from the needs of businesses to not only execute tasks but also
``to manage the flow of work through the organization,'' and managing
workflows is the natural evolution from the monolithic applications of the 1960s
to applications that rely on external functionality in the 1990s
\cite{van_der_aalst_application_1998}. By 1995, in the presence of many workflow
tools, the Workflow Management Coalition had developed a ``standard'' definition
of workflows \cite{hollingsworth_workflow_1993}:

\begin{displayquote} A workflow is the automation of a business process, in
whole or part, during which documents, information or tasks are passed from one
participant (a resource; human or machine) to another for action, according a
set of procedural rules.  \end{displayquote}

In the early 2000s, workflow systems started finding use in scientific contexts
where process automation was required for scientific uses instead of traditional
business uses. At the time, the focus of scientific workflows also shifted to
focus primarily on data processing and managing heterogeneous infrastructure for
large ``grids'' of networked services
\cite{yu_taxonomy_2005}. Yu and Buyya define a workflow as

\begin{displayquote}. . .  a collection of tasks that are processed on distributed
resources in a well-defined order to accomplish a specific goal.\end{displayquote}

This latter definition is important because of what is missing: the human
element. For many in the grid/eScience workflows community this has become the
standard definition of a workflow and the involvement of humans results not in a
single workflow, but multiple workflows spanned by a human.  Machines or
instruments are absent from the definition as well, but in practice many modern
grid workflows are launched automatically when data ``comes off'' of instruments
because they remain the primary source of data in grid workflows (cf.~\cite{megino_panda:_2015}).

In addition to grid workflows, the scientific community started exploring
``modeling and simulation workflows'' which focus not on data flow but on the
orchestration of activities related to modeling and simulation instead,
sometimes on small local computers, but often on the largest of the world's leadership class supercomputers. Unlike grid workflows they tend to require
human interaction in one way or another.  Some of these workflows are defined in
the context of a particular way of working, such as the Automation, Data,
Environment, and Sharing model of Pizzi et~al. \cite{pizzi_aiida:_2016},
the Design-to-Analysis model of Clay et~al. \cite{clay_incorporating_2015},
or the model of Billings et~al. \cite{billings_eclipse_2017}.

Additional types of workflows in the scientific community include workflows that
process ensembles of calculations for uncertainty quantification, verification
and validation or probabilistic risk assessment \cite{montoya_apex_2016}, and
workflows used for testing software. These workflows share the property
that they are all running a very large set of coordinated jobs that only provide
value when run together. However, they differ because testing workflows
typically run each test as an independent task, whereas the other workflows may
or may not change the tasks that are executed based on the intermediate state of
the entire ensemble. These workflows require a large cluster or possibly a
supercomputer in extreme cases.

Many scientific workflows have been hard-coded into dedicated environments---not
general purpose workflow management systems---that serve as point solutions
developed for the sole purpose of that single well-defined workflow, or at most
a few, to meet the needs of a single community. This leads to an important
defining characteristic for workflow management systems versus the point
solutions: workflow management systems are extensible through a public
application programming interface (API) or other method and extension does not, in
general, require the intervention of the original author. Embedding workflows
into point-solutions may be the best solution in many cases, but the
distinction between point-solutions and full workflow management systems is
important because it clearly demonstrates  that some parties prefer to focus on
rapidly creating new or modifying old workflows, whereas others may only be
interested in executing well-defined, very stable workflows.

Finally, an important class of scientific workflows is the set of ``conceptual
workflows'' that broadly define activities based on the policies of a given
community. These are common in large collaborations such as the Community Earth
System Model \cite{noauthor_cesm_nodate}. These workflows describe a
series of activities that contain both human- and computer-controlled tasks
and look like business workflows. However, depending on the author, the level of detail tends to
oscillate between very high and very low, as does the degree of abstraction. These workflows are important because they are often
referred to in the same discussions as the other types of workflows described
above. This illustrates the important fact that not all scientific workflows are
machine-executable, and it may be impossible to automate them in a workflow
management system, even one that is very good at defining abstract workflows. It
also demonstrates the difficulties that can arise in a discussion about workflows
because of ambiguity in the definition.

\subsection{Taxonomies and Classification}\label{taxonomies-and-classification}
There have been several efforts to classify, survey, or develop taxonomies for
workflows and workflow management systems, and these efforts are significant in
large part because they represent a collective call for higher order concepts in
the space. Yu and Buyya present an exceptional taxonomy for grid
workflows. Several other efforts provide highly useful vocabularies and
analyses as well.

Yu and Buyya developed a taxonomy for workflow management systems on grids that
sought to capture the architectural style and identify comparison criteria
\cite{yu_taxonomy_2005}. Their work is notable because it largely avoids a
discussion of applications and focuses purely on the functional properties of
the workflow management systems as they exist on the grids. Their work also
shows how 13 common grid workflow management systems, including Pegasus
and Kepler, fit into the taxonomy. Like other authors, Yu and Buyya cite
the lack of standardized workflow syntax and language as sources of
interoperability issues.

Scientific workflow management systems have flourished since their inception,
although not without significant overlap and duplication of effort. The survey
of scientific workflow management systems by Barker and Hemert illustrates both
growth and growing pains but also provides important observations and
recommendations on the topic \cite{barker_scientific_2007}.

Barker and Hemert also provide key insights into the history of workflow
management systems as an important part of business automation. The authors make
an important comparison between traditional business workflow management systems
and their scientific counterparts, citing in particular that traditional
business workflow tools employ the wrong abstraction for scientists. They define
workflows using the ``standard'' definition from the Workflow Management
Coalition (cf.~\S \ref{workflows} above).

The discussion points that Barker and Hemert raise are important because of
their continuing importance and relevance today, particularly the need to enable
programmability through standard languages instead of custom proprietary
languages. Sticking to standards is important and perhaps illustrated best
by Barker's and Hemert's statement:

\begin{displayquote} If software development and tool support terminates on one
proprietary framework, workflows will need to be re-implemented from scratch.
\end{displayquote}

This is an important point even for workflow tools that do not use proprietary
standards but develop their own solutions. What can be done to support those
tools and reproduce those workflows once support for continued development ends?

Montoya et~al. discuss workflow needs for the Alliance for Application
Performance at Extreme Scale (APEX) \cite{nersc_apex_2016}, and describe three
main classes of workflows: simulation science, uncertainty quantification,
and high throughput computing (HTC) \cite{montoya_apex_2016}.  HTC workflows
start with the collection of data from experiments that is in turn transported
to large compute facilities for processing. Many grid workflows are HTC
workflows, but not all HTC workflows are grid workflows since some HTC workflows---such as those presented by Montoya et~al.---may be run on large
resources that are not traditionally ``grid machines.'' When Montoya et~al.
describe scientific workflows, they are refering to the modeling and simulation
workflows described above. Montoya et~al. also provide a detailed mapping of
each workflow type to optimal hardware resources for the APEX program.

The US Department of Energy (DOE) sponsored the \emph{DOE NGNS/CS Scientific
Workflows Workshop} on April 20--21, 2015. In the report, Deelman et~al.
describe the requirements and research directions for scientific workflows for
the exascale environment \cite{deelman_future_2015}\cite{deelman_future_2017}.
The report and paper describes scientific workflows primarily by three
application types: simulations, instruments, and collaborations. The findings of
the workshop are comprehensive and encouraging, with recommendations for
research priorities in application requirements, hardware systems, system
software, workflow management system design and execution, programming and
usability, provenance capture, validation, and workflow science.

The definitions of a workflow and workflow management systems are
thoroughly explored and put into context for the purposes of the workshop. The
authors of the report are very careful to define workflows not just as a
collection of managed processes, which is common, but in such a way that it is
clear that reproducibility, mobility, and some degree of generality are required
by both the description of the workflow and the management system. (n.b. The
report appears to provide three separate definitions for ``workflow'' on pages
6, 9, and 10.)

In Reference~\cite{atkinson-csur} Atkinson et~al. discuss how to make in~silico experiments more manageable by modeling them as workflows, and to use a
workflow management system to organize their execution. They attribute the
four primary challenges of workflow execution to (i) the complexity and
diversity of applications; (ii) the diversity of analysis goals; (iii) the
heterogeneity of computing platforms, and (iv) the volume and distribution of
data. They also propose a taxonomy of workflow management system
characteristics.

Ferreira da~Silva et~al. attempt to characterize workflow management systems in
\cite{ferreira_da_silva_characterization_nodate}. The authors reduce key
properties of workflow systems into four incongruent areas: (i) design, (ii)
execution and monitoring, (iii) reusability, and (iv) collaboration. These
properties are essential considerations for most  software with limited
specificity for workflow management systems. Furthermore, there is general
conflation between classification and taxonomy and significant incoherence
between entries in equivalence classes. Most significantly, it fluctuates
somewhat chaotically between discussing workflows and workflow management
systems without linking workflow properties to the successful design and properties
of workflow systems.

\section{Experience of a Leadership Computing Facility}\label{olcf}

\subsection{Proliferation and Common Functionality} \label{commonFunc}

The problems with the increase in the number of existing workflow management
systems have been illustrated well by reports and discussions surrounding the
future of workflow management in the leadership computing facilities. The
proliferation of workflow management systems and lack of a consistent
definition of a workflow are significant barriers to the adoption of this
technology in these facilities. The High Performance Computing Facility
Operational Assessment 2015: Oak Ridge Leadership Computing Facility (OLCF)
report \cite{barker_scientific_2007} describes the problem that such
facilities face.  \begin{displayquote} These discussions concluded with the
observation that the current proliferation of workflow systems in response to
perceived domain-specific needs of scientific workflows makes it difficult to
choose a site-wide operational workflow manager, particularly for
leadership-class machines. However, there are opportunities where facilities
can centralize workflow technology offerings to reduce anticipated
fragmentation. This is especially true if a facility attempts to develop,
deploy, and operate each and every workflow solution requested by the user
community. Through these evaluations, the OLCF seeks to identify interesting
intersections that are of the most value to OLCF stakeholders.
\end{displayquote}  OLCF's strategy is notable because it makes a
very practical observation that the problem of proliferation can be solved by
consolidation of common functionality. This is typical of an operational
perspective where deployment of capability is more important than in-depth
investigation and research into how that capability functions.

\subsection{Interoperability}\label{interop}

There have been a number of community calls for interoperability. For example,
Session IV of the Twentieth Anniversary Meeting of the SOS Workshop (SOS20)
focused on workflow and workflow management system development activities of
the three participating institutions: Sandia National Laboratory, Oak Ridge
National Laboratory, and the Swiss National Supercomputing Centre
\cite{pack_sos20_2016}. Multiple presenters illustrated the challenges facing
the workflow science community and widely agreed that no single workflow
management system could satisfy all the needs of those present. Instead,
attendees proposed that the community as a whole would be served best by
seeking to enable interoperability where possible.

Workflow interoperability is not just a conceptual attribute, but one with
important practical implications. For example, DOE Leadership Computing
Facilities, as in \S\ref{commonFunc}, are affected by the lack of
interoperabilty of all types. Consider the possibility that every facility
may end up supporting different workflows systems entirely, so that workflows
at one facility can not be run at another without significant work to install
one or more additional workflow management systems! This idea is also
illustrated well in The Future of Scientific Workflows report through the
concept of the ``large-scale science campaign'' \cite{deelman_future_2015}.
Such a campaign integrates multiple workflows, not necessarily all in the same
workflow management system or at the same facility, to perform data
acquisition from experimental equipment, modeling and analysis with
supercomputers, and data analysis with either grid computing or
supercomputers.

\section{Challenges of Workflow Management Systems}\label{commonFunc}

The review of different workflow models and management systems in
\S\ref{workflows} illustrates the diversity of solutions, the lack of a coherent
understanding of workflows per se, and the absence of a coordinated search for
higher level concepts in spite of very good past efforts. That is, there is no
standard model that describes what a workflow is, the common elements of
workflow managements systems, or the description of how the pieces of such a
system interact to execute a workflow. Furthermore, there are few examples of
interoperability among existing systems in spite of significant community
pressure and calls for cross-system workflow execution. Poor or non-existent
interoperability is almost certainly a consequence of the ``Wild West'' state of
the field.

The state of the field does not mean that there is little or no common
functionality between workflow management systems in different domains. Many
sources in the literature, including several cited above, indicate
that the contrary is in fact true: there is significant duplication and
commonality in this space. The overlap in these technologies is rarely discussed
on its own merits, but instead it is commonly used to create large tables
comparing different systems, as in
\cite{ferreira_da_silva_characterization_nodate}. This creates a scenario where
more effort is spent discussing \textit{how} something is accomplished
versus the arguably more important question of \textit{what} must be
accomplished. 

Expanding on the concept of what must be accomplished, some primary application (workflow) needs include
(i) lowering the development burden; (ii) extensibility; (iii) transporting an
application workflow to another resource, platform, or workflow system; and
(iv) providing a conceptual framework or basis to decide which tools are suitable or optimal for a given workflow.

Similarly, beyond having clarity on the functional and performance capabilities of a
workflow system, the primary needs of users and developers of workflow systems include (i) lowering the need to develop components, (ii) determining which components
to use and reuse, (iii) minimal perturbation and refactoring when extending or
generalizing the functionality or use cases supported by a workflow system,
(iv) providing constant performance across different use-case scenarios and
scales.

It is worth noting that workflow systems are rarely developed to extract
(enhance) performance. They are more about coordinating different
functionality without loss of performance. High-performance and scalability is
not often a first order concern of general workflow systems; it may however,
be a first order concern of specialized workflow systems or specific
components (e.g, a pilot-system that is responsible for scalable and efficient task launching and management).

A healthy balance of what versus how is
important, but we propose that the discussion of how particular problems are
solved in workflow science has overtaken the discussion of what must be
accomplished, creating two severe problems: \begin{itemize} \item A
``proliferation'' of tools that largely solve the same problem in the same way,
but with separate, competing implementations primarily delineated along domain,
as opposed to technological, boundaries.  \item A general lack of
interoperability and therefore inability to address larger scientific problems
using hybrid combined workflows, multifacility workflow campaigns, or
heterogeneous hardware without significant reimplementation.  \end{itemize}

These two problems are closely related: Tooling proliferation might not be a
problem, given sufficient resources, in the absence of calls for
interoperability between systems, and interoperability might not be an issue if
there were not so many existing systems. However, some of the most important
aspects of these problems remain separable and should be examined as such.

Workflow interoperability is neither a simple nor singular attribute. There
are at least four distinct types of interoperability that merit discussion:

\begin{enumerate}

\item Workflow interoperability---Sharing workflows across different science
problems. This was an original motivation in the initial days of eScience  and
reproducible computational science. Early projects such as the MyGrid
(subsequently MyExperiment) and related projects, pioneered and advanced the
ability to share workflows across science domains, science  problems and
scientists.

\item Execution delegation---Delegating the execution of a workflow to a more
capable or appropriate workflow management system. Consider, for example, the formal specification of a workflow as a directed acyclic graph and associated data descriptions, such that the specification is complete and thereby in principle executable by any capable workflow management system.  Although in principle and
conceptually easy, this has proven to be less successful in practice for at least
two primary reasons: (i) directed acyclic graphs are a common, but not universal, formal specification of some workflows, and (ii) many specific consideration and assumptions beyond those
associated with a directed acyclic graph need to be factored when executing workflows. These
assumptions and specific considerations in turn are often due to inadequate
infrastructure abstraction and separation of concerns. 

\item Workflow system interoperability---Executing the same workflow(s) by
different workflow management systems. In addition to the absence of a
technical or formal basis for designing workflow management systems, the
sociology of software engineering and tooling contributed to the proliferation
of workflow management systems. In the presence of a proliferation of tools,
there was always a principled if not a practical demand for such workflow
system interoperability. However, even if initially a  more  ``principled'' form
of interoperability, it can be argued that workflow system interoperabilty is
increasingly important because of the needs and requirements of reproducible
science.

\item Interchangeable workflow system components---Components that can be
exchanged or used concurrently across one or more systems. Until now, this is
the least articulated or argued form of interoperability. However, it is
the most critical and core form of interoperability that our work suggests 
must be addressed, if the component based approach to workflow systems is
ever to supplement monolithic workflow systems.

\end{enumerate}

A primary driver for seeking interoperability across workflows systems
has been the need to address larger scientific problems that can only be
solved with workflows that require multiple systems for complete execution.
Two successful examples of limited interoperability between workflow systems
are discussed in \cite{brooks_triquetrum:_2015} and
\cite{mandal_integrating_2007}. Notably both projects leveraged flavors of
the Ptolemy framework, namely Triquetrum and Kepler, and delegated the
execution of workflows.

\section{The Solution: Common Building Blocks}\label{buildings-blocks}

The two problems detailed above are side effects of the relentless march of
progress. The traditional approach for building workflow systems has been to
build as much of the required capability as possible into the system itself,
relying very little on external services or even third party code to address
pressing issues in one or more domains. However, history has shown that
important high-level functionality slowly moves down the software stack and
into kernels, kernel services, and system libraries. Is it better at that
point to use an existing system that requires significant time and resources
to learn, or to develop yet another workflow management system with common
tools, implementing only the gaps instead?

The answer to this question is complicated by the fact that workflows
themselves have evolved. First, contemporary workflows are often the
representation of methodological advances and may be more pervasive,  short-
lived, and wide-ranging than traditional workflows. Further, they are no longer
confined to ``big science'' projects because sophisticated workflows are needed
by many types of scientific projects, which leads to diverse design features and thus
makes it unlikely that one model will be universally applicable.  The ability to prototype, test
and experiment with workflows at scale suggests a need for interfaces and
middleware services that enable the rapid development of resources. The
challenge is to provide these capabilities along with considerations of
usability and extensibility.

Jha and Turilli discuss this trend as it relates to workflows from a 
cyber-infrastructure perspective and to existing large-scale scientific workflow
efforts \cite{jha_building_2016}. They propose that, while historically
successful, monolithic workflow systems present many problems for users,
developers, and maintainers. Instead, they propose that a new ``Lego-style''
approach might work better where individual building blocks of capability
are assembled into the final workflow management system, subsystem, or
product.

More formally, a building block is a collection of functionality commonly
identified across existing workflow systems that behaves like a logically and
uniformly addressable service. Table \ref{blocks} lists six common types of
functionality that are readily observed in workflow management systems. There
are certainly additional types of functionality that are common, but for
pedagogical reasons we limit the list to the most obvious choices in a quick review of the literature previously cited.

\begin{table*}[h]
\begin{tabu} to \textwidth {|X[l]|X[l]|} \hline
\textbf{Functionality} & \textbf{Description} \tabularnewline\hline Data and
metadata management & Management of data, metadata, and general file input and
output activities whether for internal tracking or external user
consumption. \\ \hline 
Workflow execution engine & The primary
actor that manages the execution of the activities as provided by the workflow
description. \tabularnewline\hline 
Resource management and acquisition &
Acquisition and management of resources, whether computing or instrumentation,
required for the successful execution of the workflow. \tabularnewline\hline
Task management & Primary subsystem for managing individual activities, tasks or
``subworkflow'' using resources provided by the task management system. This
system is sometimes, but neither often nor exclusively, part of the workflow
execution engine. \tabularnewline\hline 
Provenance engine & System for tracking
execution history, sources, and destinations of ingested and generated artifacts,
execution metadata including status, general logging, and provenance-based
inference tools. \tabularnewline\hline 
Application programming interface (API) &
A non-functional element of most workflow management systems that is critical to
successful deployment and maintenance of the full system as well as use
as a tool for creating and executing workflows. \tabularnewline\hline
\end{tabu} 
\caption{Functionality commonly identified in workflow management
systems.} 
\label{blocks}
\end{table*}

Each of the types of functionality listed in Table
\ref{blocks} could be developed, presumably through one or more community
efforts, as a building block (even the API through some programming
trickery!). Other things like programming interfaces to queuing systems,
programmable pilot systems for scheduling jobs, workload balancers, and
ensemble execution tools, among others, could be provided as well to create a
rich ecosystem of reusable and interchangeable parts.

Reusable building blocks would greatly improve both interoperability and
sustainability because they would standardize, to some degree, the programming
interfaces and back-ends used by workflow management systems. To the extent that
projects are willing to use common building blocks, proliferation would be fully
decoupled from interoperability. Leadership computing facilities would not need
to support every workflow management system, just a set of common building
blocks. This is similar to how they support third-party libraries for software
development: they do not support every code used on these machines, but
they support a set of common libraries that the codes can use. 

There is an important practical question here: Does this mean abandoning
existing workflow management systems or redeveloping existing workflows? No, and
in fact it may be quite practical to develop building blocks based on components
of the most sophisticated workflow management systems already in existence.
Furthermore, because building blocks would naturally enable interoperability, it
is quite conceivable that a workflow that only executes on one system now may
execute on many systems in the future with little or no modification. 

A second question is whether or not building blocks represent a significantly new
type of modularity versus a traditional software stack or framework. Building blocks
arguably sit above these entities and have distinct conceptual and functional roles.
A software stack is the full set of software, including all dependencies, for a given
application or software product and a framework is the set of common functionality
(APIs, not libraries) around which the product is built. On the other hand, a
building block may be implemented using a framework and will have some software stack,
but it will also offer a complete set of functionality that can be used directly in
an application. The building block may also be offered on a different system with a
different implementation (i.e., using a different software stack and framework), but neither its functionality nor service interface would change.
\section{Discussion and The Road Ahead}\label{discussion}

This paper is about the practice of using workflow systems in general, as
opposed to the experience of a specific workflow system. It is motivated by
the widely shared perception, if not strong empirical evidence and observation
that there is a problem in the current practice of workflow management
systems. The paper describes a variety of problems and challenges commonly
found in the workflow science space. Self-evidently, no single workflow
management system will be able to address the next generation of scientific
challenges and practical experience dictates that a change is necessary.

There is an important separation between the challenges of expressing
workflows effectively versus a workflow system that will execute the workflow.
In this paper, we do not discuss the challenges inherent in expressing
workflows effectively. Further, this work is not a theoretically motivated or
survey paper about models of workflows or workflow systems; although plenty of
such papers exist, their impact on the practice of workflow systems has been
limited.

It is illustrative if not instructive to understand the ecosystem of the
Apache BigData Software Stack/Cloud Model, where there are many seemingly
similar components for  data-intensive workflows. The proliferation of
components suggests there is an strong preference of functional specialization
and diversity of use, as opposed to interoperability.  Equivalentally, there
is a strong binding of components to platforms.

In response to the problems and experience, we propose that common components
in the form of building blocks are a promising and practical solution. We
suggest that a building blocks approach will solve problems of system
proliferation and interoperability by harnessing and developing common
functionality that exists in workflow management systems into reusable
services.

An important and critical test will be to devise a validation (or negation)
test for the hypothesis that a building blocks approach to workflows is in fact
more scalable, sustainable and better practice than monolithic workflow
systems. We do not harbor illusions that it will be easy, or that it is
necessarily even possible.

\begin{acks}

The authors are grateful for the assistance and
support of the following people and institutions without which this work would
not have been possible.

Mr. Billings is especially grateful for the feedback provided on
\S\ref{workflows} and \S\ref{commonFunc} by his PhD committee, including Jack Dongarra, John Drake, Mike Guidry, Mallikarjun Shankar, and John Turner. Mr.
Billings would also like to acknowledge the thoughtful discussions with Jim
Belak on the nature of workflows in the ExAM project, and Robert Clay, Dan
Laney, and David Montoya on modeling and simulation workflows.

This work has been supported by the US Department of Energy, the Oak Ridge National Laboratory (ORNL) Director's Research and Development Fund, and by the ORNL
Undergraduate Research Participation Program, which is sponsored by ORNL and
administered jointly by ORNL and the Oak Ridge Institute for Science and
Education (ORISE). ORNL is managed by UT-Battelle, LLC, for the US Department
of Energy under contract no. DE-AC05-00OR22725. ORISE is managed by Oak Ridge
Associated Universities for the US Department of Energy under contract no.
DE-AC05-00OR22750.

\end{acks}

\bibliographystyle{ACM-Reference-Format}
\bibliography{bib}

\end{document}